\documentclass[10pt,aps,prb,twocolumn,superscriptaddress,floatfix,showpacs,longbibliography]{revtex4-1}
\usepackage[utf8]{inputenc}
\usepackage{graphicx}
\usepackage{tabularx}
\usepackage[usenames,dvipsnames,table]{xcolor}
\usepackage[version=3]{mhchem}
\usepackage{braket}
\usepackage{upgreek}
\usepackage{hyperref}
\usepackage{amsmath, amssymb}
\usepackage[normalem]{ulem}
\usepackage{soul}

\definecolor{mark}{rgb}{0.85, 0.9, 1}
\definecolor{rred}{HTML}{CB4154}

\hypersetup{hidelinks}
\sethlcolor{mark}
\newcolumntype{Y}{>{\centering\arraybackslash}X}

\begin{document}

\title{Reconstruction of classical skyrmions from Anderson towers: \\
quantum Darwinism in action}
\author{O.~M.~Sotnikov$^1$, E.~A.~Stepanov$^2$,  M.~I.~Katsnelson$^3$, F.~Mila$^4$, V.~V.~Mazurenko}

\affiliation{Theoretical Physics and Applied Mathematics Department, Ural Federal University, Mira Str. 19, 620002 Ekaterinburg, Russia\\
$^2$ CPHT, CNRS, Ecole Polytechnique, Institut Polytechnique de Paris, F-91128 Palaiseau, France\\
$^3$Radboud University, Institute for Molecules and Materials, Heyendaalseweg 135, 6525AJ, Nijmegen, Netherlands\\
$^4$Institute of Physics, \'Ecole Polytechnique F\'ed\'erale de Lausanne (EPFL), CH-1015 Lausanne, Switzerland}
\date{\today}

\begin{abstract}
The development of the quantum skyrmion concept is aimed at expanding the scope of the fundamental research and practical applications for classical topologically-protected magnetic textures, and potentially paves the way for creating new quantum technologies. 
Undoubtedly, this calls for establishing a connection between a classical skyrmion and its quantum counterpart: a skyrmion wave function is an intrinsically more complex object than a non-collinear configuration of classical spins representing the classical skyrmion. 
Up to date, such a quantum-classical relation was only established on the level of different physical observables, but not for classical and quantum states \emph{per se}. 
In this work, we show that the classical skyrmion spin order can be reconstructed using only the low-energy part of the spectrum of the corresponding quantum spin Hamiltonian.
This can be done by means of a flexible symmetry-free numerical realization of Anderson's idea of the towers of states (TOS) that allows one to study known, as well as unknown, classical spin configurations with a proper choice of the loss function. We show that the existence of the TOS in the spectrum of the quantum systems does not guarantee \emph{a priori} that the classical skyrmion magnetization profile can be obtained as an outcome of the actual measurement. This procedure should be complemented by a proper decoherence mechanism due to the interaction with the environment. The later selects a specific combination of the TOS eigenfunctions before the measurement and, thus, ensures the transition from a highly-entangled quantum skyrmionic state to a classical non-collinear magnetic order that is measured in real experiments. The results obtained in the context of skyrmions allow us to take a fresh look at the problem of quantum antiferromagnetism. In particular, we provide a quantitative characterization of the TOS contributions to classical antiferromagnetic structures including frustrated ones.   
\end{abstract}

\maketitle

\section*{Introduction}
Despite the huge success of quantum physics, which is one of the pillars of our science and technology, its foundations still remain a subject of hot debates, and many conceptual issues still require further investigations~\cite{qmi1,qmi2,qmi3,qmi4,qmi5,qmi6,qmi7}. 
In particular, searching for the connection between quantum and classical descriptions of the same phenomenon or object has a long history in physics starting from the foundation of quantum mechanics. 
In this sense, the development of the path integral concept~\cite{Feymann1,Feymann2, Feymann3} is a bright example showing that the classical trajectory of a particle is just one of numerous alternatives characterized by different probabilities. 
In these terms, classicality means nothing but destruction of an interference between different alternatives, similar to a transition from wave to geometric optics~\cite{born_wolf}. 

However, in the case of the classical-quantum correspondence the problem is more complicated. The transition between the classical and quantum regimes cannot be determined only by the fact that the characteristic size of the system, which can be related to the de Broglie wavelength, becomes small compared to other length scales of the problem. 
According to the popular decoherence program, the classical-quantum correspondence is rather related to the openness of the quantum system and to the destruction of quantum interferences by the interaction with the environment~\cite{Joos2,Joos1,Zurek0,Zurek1,Zurek3,Zurek2}. 
The problem is closely connected to the measurement problem~\cite{Wheeler_Zurek}. 
According to Bohr's complementarity principle~\cite{Bohr}, a quantum measurement is nothing but the result of the interaction of a quantum particle with a classical measuring device. 
This picture is the basis of the formal theory of measurements developed by von Neumann~\cite{neumann}. 
This theory includes a mysterious collapse of the quantum wave function after the measurement. 
Further developments have led to a more complicated picture, including soft measurements~\cite{Mensky} and decoherence waves in distributed quantum systems~\cite{decoherence_waves1, decoherence_waves3, decoherence_waves2}. 
There are also analytical~\cite{theo} and numerical~\cite{hylke} attempts to derive von Neumann's postulate from a consequent quantum consideration of the measurement process, including decoherence by the environment. 
In general, the problem does not seem to be completely solved, and further attempts at clarifying these key issues are required. 

These questions may look too general and too abstract but, actually, they are very closely related to a very common and important phenomenon of physics around us. 
Antiferromagnetism, a very usual property of condensed matter~\cite{vonsovsky}, is, probably, one of the best examples.
The classical N\'eel picture of magnetic sublattices for the case of an ``antiferromagnetic'' exchange interaction is in an obvious contradiction with quantum mechanics predicting a singlet ground state~\cite{vonsovsky}. 
Actually, the antiferromagnetic state can be described without introducing sublattices~\cite{irkhin}, but its difference with the singlet state remains dramatic. 
A general way to establish the correspondence between quantum and classical descriptions of antiferromagnets was open in the seminal work of P.~W.~Anderson~\cite{Anderson}. It has been shown there that in some cases linear combinations of eigenstates of a quantum Hamiltonian that form a tower of low-energy states can be related to an ordered state that would be the classical ground state of the system in the thermodynamic limit. 
Such a tower-of-states (TOS) approach may be of fundamental importance because it treats the fundamental problem of quantum-classical correspondence from a completely different perspective, without referring to measurements or postulating decoherence due to the environment. 
Moreover, it has proven to be extremely helpful in detecting broken symmetries with eigenspectrum of even small-size supercells of quantum systems. 
Up to now, the Anderson towers approach was mainly used for studying quantum antiferromagnets~\cite{Misguich, Frederic, Bernu}. 
However, it is worth mentioning that the previous studies based on the group-theoretical calculations were fully concentrated on the symmetry identification of the eigenfunctions contributing to the TOS without attempting to quantify their partial contributions. 
Such an approach is also not flexible since it requires to know the exact symmetry of the reconstructed classical order, which prevents using the approach in the case when the system is characterized by a transition to an unknown classical state (a problem known as hidden order).  

In this paper, we report on a symmetry-free numerical technique based on gradient-descent optimization for constructing TOS on the basis of a limited number of calculated low-lying eigenstates of a quantum system. 
In contrast to previous works, our approach provides quantitative information on the TOS composition. 
By means of the developed scheme we explore topologically-protected classical magnetic skyrmions~\cite{Bogdanov} that attract a considerable attention due to their fundamental interest~\cite{Skyrmion_roadmap, DMI_guide, light} and technological importance in design of atomic ultra-dense memory~\cite{pureDMI}, probabilistic computing~\cite{skyrmion_random_numbers}, etc. 
Up to now, most of the studies~\cite{Skyrmion_roadmap} have been focused on a pure classical description of the skyrmionic structures, which assumes that each spin in the system is a classical vector with three spatial projections. 
The experimental discovery of nano-scale skyrmionic structures \cite{Wiesendanger} in surface nanosystems significantly heats up interest to search for quantum analogues of classical skyrmions, for which quantum effects may play a crucial role~\cite{Fernandez}.

Recent theoretical studies~\cite{qsk1,qsk2,qsk3,qsk4,qsk5,qsk6, Quantum_skyrmion} suggested that the ground state of some quantum spin Hamiltonians with competing isotropic and anisotropic interactions can be considered as analogs of classical skyrmions since the magnetization, the susceptibility, and the scalar chirality calculated for these quantum ground states agree with those obtained for the corresponding classical models. 
However, as was shown in a previous work~[\onlinecite{Quantum_skyrmion}], both projective measurements and site-resolved average magnetization calculated in the ground state of periodic quantum systems do not resemble the typical pattern of a classical skyrmion as seen in experiments. 
For this reason, one can formally define the quantum skyrmion as a quantum state for which the spin-spin correlation functions reproduce the same quantities in the classical version of the problem. 
The natural question is then what is the mechanism through which one can observe a classical skyrmion in a system which is a priori quantum. 

In this work, we show that the connection between the classical and quantum skyrmion systems can be established not only at the level of the observables, as done up to now, but also at the more general level of a quantum state and of macroscopic classical order \emph{per se}. 
To this aim we use the concept of Anderson's tower of states and explore both the towers of quantum wave functions to reconstruct classical solutions, and the towers of classical configurations needed to reproduce the quantum ground state. 
The analysis of the composition of the TOS that we have obtained for magnetic skyrmions and for antiferromagnets uncovers important details of Anderson's theory that were not addressed in previous group-theoretical considerations. 
We argue that the environment plays a crucial role in selecting specific combinations of the quantum states that build the TOS that correspond to classical order and are known as pointer states in decoherence theory (Quantum Darwinism). Thus, quantum decoherence should be considered as an important part of the TOS theory. Since the TOS analysis itself can be used as a more effective search for pointer states within the decoherence program, both theories will benefit from this integration. 

\begin{figure*}[!t]
	\includegraphics[width=1.8\columnwidth]{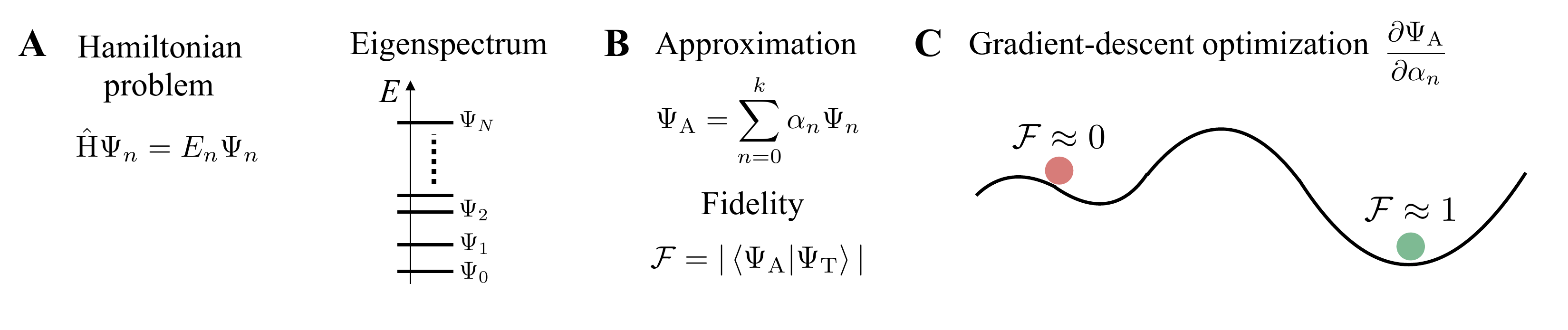}
	\caption{\label{intro} Protocol for constructing Anderson's towers of states. (A) For the given Hamiltonian one calculates a set of low-lying eigenstates. (B) On the basis of the calculated eigenstates, an initial approximation of the target state is prepared. The complex coefficients $\alpha_n$ are chosen to be random.  (C) The coefficients are optimized within a gradient-descent approach aiming to maximize the fidelity between approximation and target wave functions. $N$ is the corresponding number of energy levels.}
\end{figure*}

\section*{Results}
\subsection*{Protocol for constructing the Anderson tower}
Formally, a quantum wave function corresponding to any classical spin texture can be defined as a product of coherent states of individual spins \cite{Moskvin, coherent_state, inomata}
\begin{eqnarray}
\label{target}
\ket{\Psi_{\rm T}} = \prod_{i} [ {\rm cos} \frac{\theta_i}{2} e^{i\frac{\phi_i}{2}} \ket{\uparrow} + {\rm sin} \frac{\theta_i}{2} e^{-i\frac{\phi_i}{2}}\ket{\downarrow}],
\end{eqnarray}
where the polar angles $\theta_{j}$ and $\phi_{j}$ set a local basis for each spin. 
Below we will refer to this state as a coherent state or a target wave function. 
The consequent projective measurements~\cite{neumann} of the state $\Psi_{\rm T}$ in $\sigma^z$, $\sigma^x$ and $\sigma^y$ bases result in a set of projections $\langle \hat{S}_i^z \rangle$, $\langle \hat{S}_i^x \rangle$, and $\langle \hat{S}_i^y \rangle$ for each spin. 
The latter can be associated with the direction of the classical magnetic moment ${\bf m}_{i}$ in magnetic structures measured in spin-polarized scanning tunneling microscopy experiments~\cite{Wiesendanger}. 
More specifically, ${\langle m^x_i \rangle = \sin \theta_i \cos \phi_i}$, ${\langle m^y_i \rangle = \sin \theta_i \sin \phi_i}$, and ${\langle m_{i}^z \rangle = \cos \theta_i}$. 
Thus, one can establish a formal connection between parameters of the coherent state and the classical magnetic moments of a quantum system observed in real or numerical experiments: ${\theta_{i} = \arccos \langle m_{i}^z \rangle}$ and ${\phi_i = \arctan \frac{\langle m^y_i \rangle}{\langle m^x_i \rangle}}$.   

In order to construct the Anderson tower for a quantum system we follow the key steps visualized in Fig.~\ref{intro}\,\mbox{A-C}. First, we perform the exact diagonalization of a quantum Hamiltonian and determine its eigenstates $\Psi_n$ (Fig.~\ref{intro}\,A). We consider only the low-lying part of the eigenspectrum ${n\in[0,k]}$ and introduce the initial approximation for the target wave function $\Psi_{\rm A}$ with random complex coefficients $\alpha_n$  (Fig.~\ref{intro}\,B). 
Further, these coefficients are varied using the gradient-descent method to get the maximal fidelity between $\Psi_{\rm A}$ and $\Psi_{\rm T}$ (Fig.~\ref{intro}\,C). 
Here, as for any optimization procedure, the choice of the loss function that is responsible for the quality of the resulting approximation and convergence speed plays a central role. In this work it is given by the following expression:
\begin{eqnarray}
\mathbb{E} (\mbox{\boldmath$\alpha$}) = 1- |\braket{\Psi_{\rm T}|\Psi_{\rm A} (\mbox{\boldmath$\alpha$})} |. 
\label{eq:loss}
\end{eqnarray}
The coefficients are updated as
\begin{eqnarray}
\mbox{\boldmath$\alpha$}_{\rm new} = \mbox{\boldmath$\alpha$}_{\rm old} - \gamma \frac{\partial \mathbb{E}}{\partial \mbox{\boldmath$\alpha$}_{\rm old}},  
\end{eqnarray}
where $\gamma$ is the gradient-descent step that is taken to be $1$.
This choice for the loss function can be justified by the fact that the fidelity is a standard metric to define the distance between the two quantum states~\cite{Nielsen}.

\begin{figure*}[!t]
	\includegraphics[width=1.8\columnwidth]{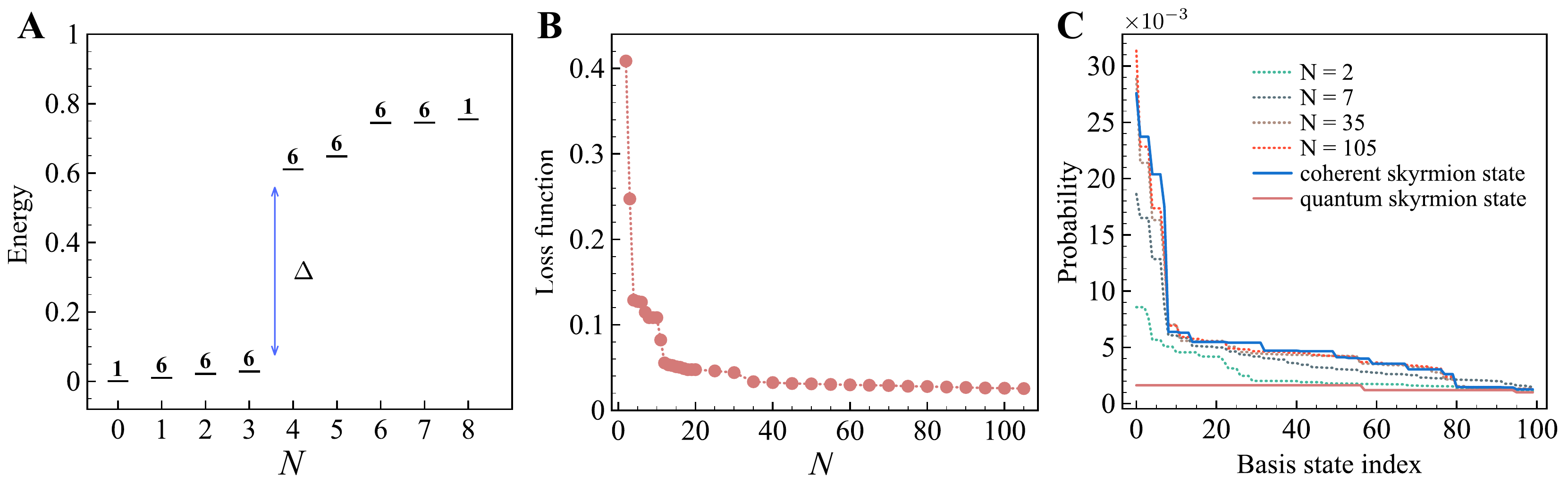}
	\caption{\label{gradient}  Structure of the eigenspectrum and performance of the gradient-descent optimization. (A) Low-energy part of the eigenspectrum of the quantum Hamiltonian, Eq.\ref{Ham} obtained with $J$ = 0.5, $|{\bf D}|$ = 1 and $B$ = 0.44. The numbers denote the degeneracy of each eigenlevel. (B) Dependence of the loss function on the number of eigenstates of minimal energy. (C) Comparison of the ordered distributions of quantum states amplitudes. Only the hundred basis functions giving the largest contribution to the particular state are considered.}
\end{figure*}

An important property of the proposed numerical scheme for constructing TOS is that we can quantitatively control the quality of the approximation for $\Psi_{\rm A}$ with the parameter $k$ that truncates the eigenspectrum.  
It is worth noting that in previous works the construction of the Anderson TOS for antiferromagnets was mainly based on a symmetry-based selection of the low-lying states of the eigenspectrum in order to find a signature of a symmetry-broken state that can be the ground state of the system in the thermodynamic limit~\cite{Wietek}.  
Such a symmetry-based realization of the Anderson idea limits its possible applications and does not provide quantitative information on the contribution of a particular eigenstate to the TOS, which, as we show below, can be done with our protocol. 
In addition, in our approach the loss function~\eqref{eq:loss} can be replaced by another form with milder conditions, which might be useful in the case where the precise form of the target state $\Psi_{\rm T}$ is unknown.

\subsection*{Skyrmionic tower of states}
The implementation of the proposed protocol in the case of magnetic skyrmions requires a specific choice of spin Hamiltonian. 
While various microscopic mechanisms for stabilizing skyrmion structures can be considered in the quantum case~\cite{qsk1,qsk2,qsk3,qsk4,qsk5,qsk6}, we focus on the most traditional one which is based on the competition between the isotropic and the anisotropic exchange interactions in the presence of an external magnetic field. The corresponding quantum Hamiltonian on a two-dimensional lattice is given by: 
\begin{align}
\hat {\rm H}_{\rm sk} = \sum_{ij} J_{ij} \hat{\bf S}_i  \cdot  \hat{\bf S}_j + \sum_{ij} {\bf D}_{ij} [\hat{\bf S}_i \times \hat{\bf S}_j] + \sum_{i} B \hat{S}^z_i.
\label{Ham}
\end{align}
Here, $J_{ij}$ is the isotropic Heisenberg exchange interaction. $\mathbf{D}_{ij}$ is an in-plane vector that points in the direction perpendicular to the bond between neighboring $i$ and $j$ lattice sites and describes the Dzyaloshinskii-Moriya interaction (DMI). $\mathbf{B}$ is an external uniform magnetic field applied along the $z$ direction. In the following, for constructing the skyrmionic towers, we use the parameters $J$ = 0.5, $|{\bf D}|$ = 1 and $B$ = 0.44, which guarantee the stabilization of the quantum skyrmion wave function as the quantum ground state~\cite{Quantum_skyrmion}.  

In this work we are interested in exploring the properties of the eigenspectrum, thus the exact diagonalization of the constructed Hamiltonian is performed. Following Ref.~[\onlinecite{Quantum_skyrmion}], a 19-site supercell with periodic boundary conditions on the triangular lattice is considered.
The corresponding technical details are described in the Methods section. 
The low-energy part of the eigenspectrum is presented in Fig.~\ref{gradient}\,A and is characterized by a sizeable gap between the first 19 eigenstates and the rest of the spectrum. 
The dependence of the gap on the value of the magnetic field is discussed in the Methods section. 
As we will show below, the states at the bottom of the gap play a central role in reconstructing the coherent skyrmion state with Anderson towers. 
According to Fig.~\ref{gradient}\,B the loss-function demonstrates a steep decrease for ${N < 20}$, where $N$ is the number of  eigenlevels taken into account. Further increasing $N$ leads to the saturation of the loss function.
Importantly, the number of eigenstates we consider is negligibly small with respect to the total dimension of the Hilbert space. More specifically, while the total dimension of the Hilbert space is equal to $2^{19}$, to construct an approximation of the coherent skyrmion state we use superpositions of up to 455 eigenstates that can be packed into 105 different eigenlevels taking into account the degeneracy of the eigenfunctions.

\begin{figure}[!t]
	\includegraphics[width=\columnwidth]{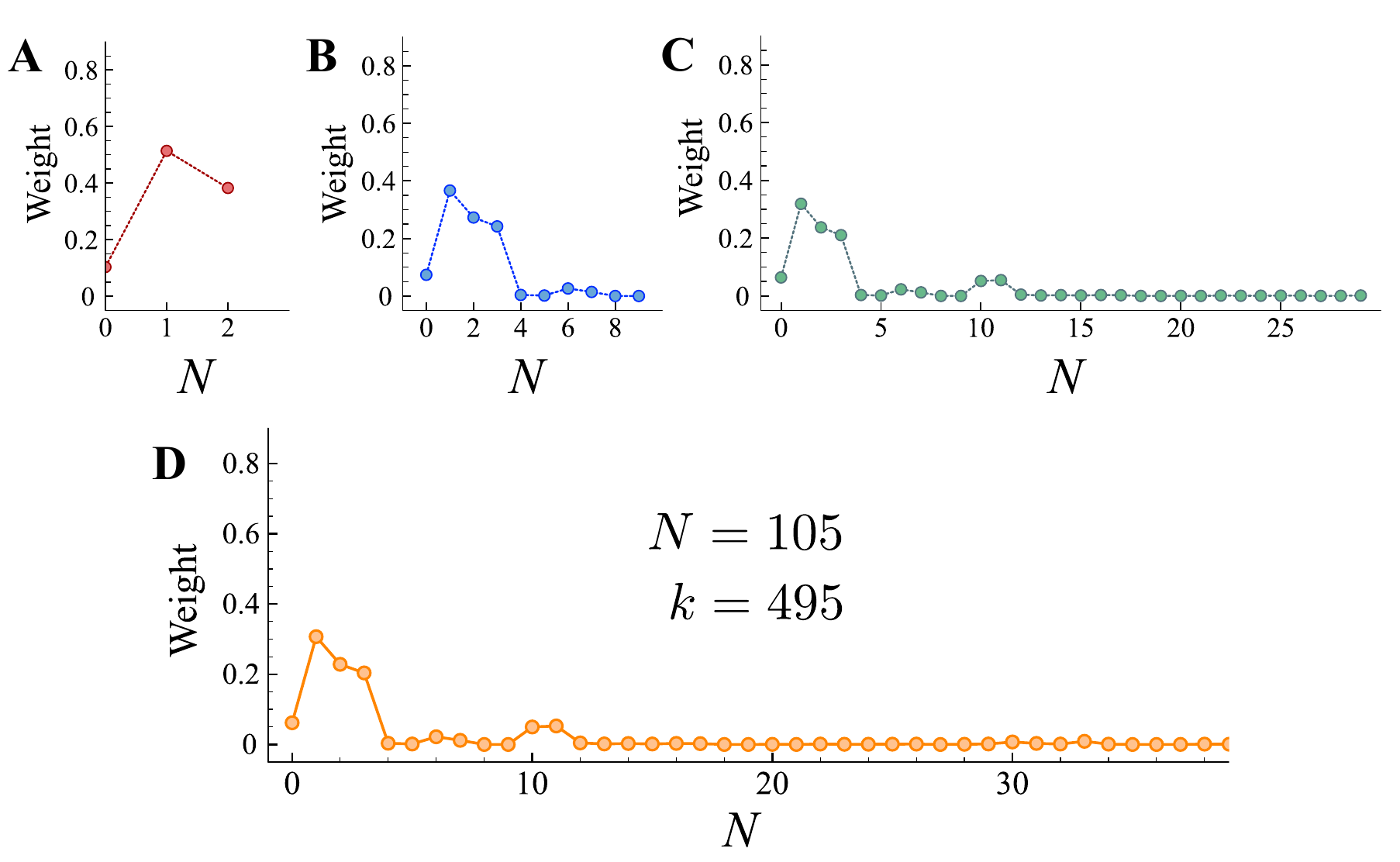}
	\caption{\label{weight} Structure of the coherent skyrmion state approximations calculated with different number of states from the low-lying part of the eigenspectrum of the Hamiltonian Eq.\ref{Ham}. The points denote the sum of probabilities ($\alpha^2_{n}$) of the eigenstates corresponding to the same energy (eigenlevel). $N$ denotes the number of eigenlevels with different energies.}
\end{figure}

\begin{figure}[!t]
	\includegraphics[width=\columnwidth]{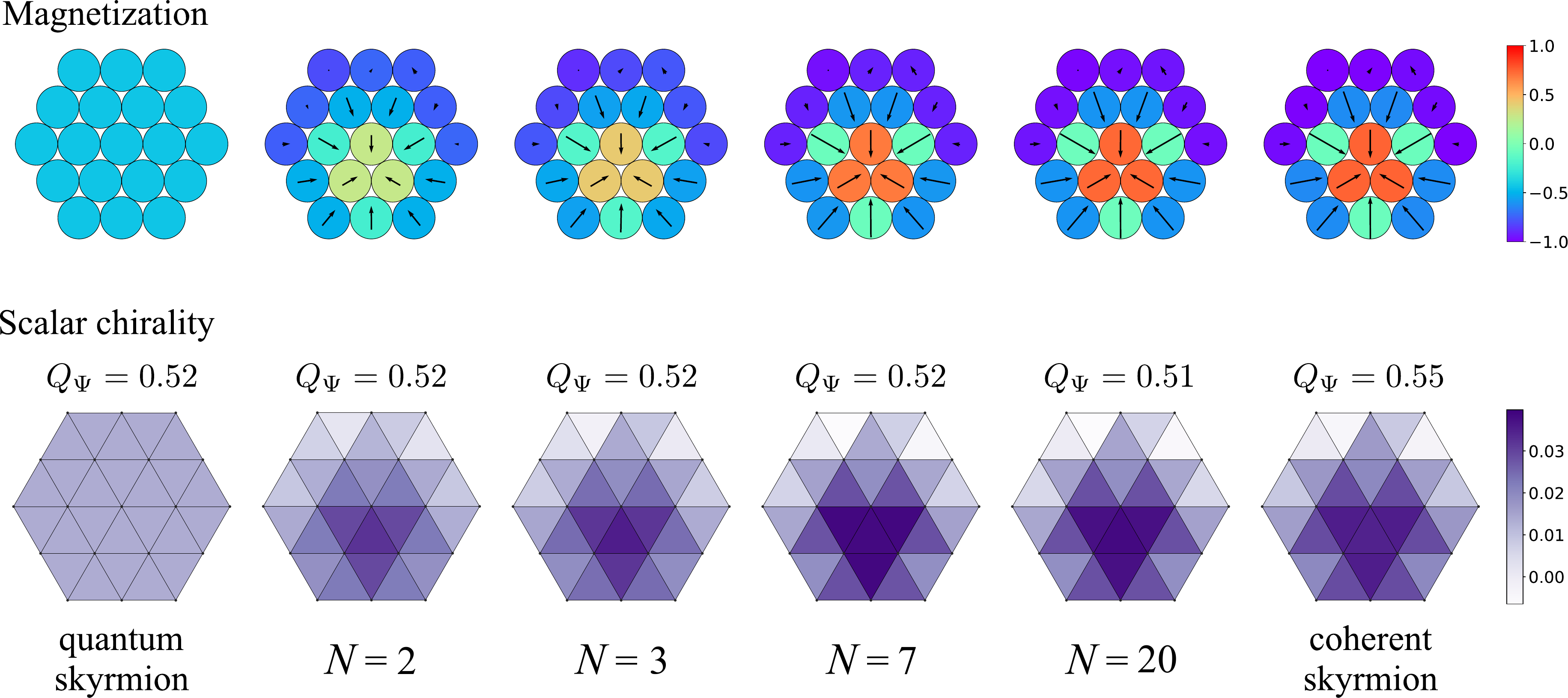}
	\caption{\label{chirality}  Magnetization (top) and scalar chirality (bottom) calculated for the reconstructed wave function corresponding to the classical skyrmion with different number of eigenstates of the parent spin Hamiltonian. The arrows denote the in-plane direction of spins while the $z$ components are defined with color. The triangle-resolved scalar chirality $Q_{ijl}^{\Psi}$ denoted with shaded triangles is calculated for real quantum skyrmion state (left plaquette), coherent state with classical profile of the magnetization (right plaquette) and for quantum states that are different approximations of the classical skyrmion within the Anderson tower approach. $N$ denotes the total number of low-lying energy levels involved in the state reconstruction.}
\end{figure}

Fig.~\ref{gradient}\,C shows the comparison of the probabilities of the basis functions that give the largest contribution to the quantum skyrmion and to the coherent skyrmion state and its approximations with different numbers of $N$. 
The probability function for $\Psi_{\rm T}$ is characterized by a step-like profile, which means that this quantum state is highly-structured in the Hilbert space. 
By contrast, the quantum skyrmion is strongly delocalized. Fig.\ref{weight} shows the composition for some approximations of the coherent skyrmion state. 
The largest weights ($\sum_{n}\alpha^2_{n}$, where $n$ corresponds to a particular eigenlevel) are provided by degenerate states corresponding to the first three excited energy levels. 
The eigenstates of larger energies give much smaller contributions, but they are important to reach the high fidelity in reproducing the solution of the classical version of the Hamiltonian~\eqref{Ham}. Indeed, as can be seen from Fig.~\ref{chirality}, the local magnetization of the quantum Hamiltonian~\eqref{Ham} is uniformly distributed on the lattice.
We find that already the first approximation for $\Psi_{\rm A}$ that takes into account a small number 
$N$ of levels leads to a magnetization profile that reminds one of the classical skyrmion spin texture.
Such an approximation is formed with the ground state and the 6-fold degenerate first excited state. 
Further extension of the set of excited states increases the fidelity between the target wave function $\Psi_{\rm T}$ and its approximation $\Psi_{\rm A}$.
For ${N = 105}$ the fidelity reaches the value of 97.5\%. 

For a complete characterization of the transition from the quantum skyrmion wave function to the classical skyrmion state, the performed investigation of the magnetization pattern should be supplemented by the calculation of the scalar chirality and its distribution over the system in question~\cite{Quantum_skyrmion}. 
This quantity is given by the following expression:
\begin{eqnarray}
Q_{ijl}^{\Psi} = \frac{1}{\pi}  \langle \hat {\bf S}_{i} \cdot [\hat {\bf S}_{j} \times \hat {\bf S}_{l}]  \rangle,
\label{QPsiloc}
\end{eqnarray}
where $i$, $j$, and $l$ depict three neighboring spins that form an elementary triangular plaquette. The total chirality is defined as a sum of local chiralities of individual plaquettes 
\begin{eqnarray}
Q_{\Psi} = \sum_{\langle{ijl\rangle}} Q_{ijl}^{\Psi}. 
\label{QPsi}
\end{eqnarray}
We note that if the Hamiltonian~\eqref{Ham} is defined on an infinite lattice, the problem can be solved on the basis of a supercell with periodic boundary conditions, and all individual triangular plaquettes produce the same contribution to the total chirality. 
The distribution of the chirality is thus uniform, as the distribution of the magnetization in the system. 
As the bottom row in Fig.~\ref{chirality} shows, in the case of the classical skyrmion state $\Psi_{\rm T}$ and its different approximations $\Psi_{\rm A}$, the scalar chirality has a non-uniform distribution with the largest contributions from the triangles characterized by the strongest variation of the magnetization, which can be explained by the smallness of the system. 
Importantly, the values of the total chirality are in the range 0.52 - 0.56 within the series of approximations as well as for the target wave function $\Psi_{\rm T}$ that corresponds to the classical skyrmion structure.   

\subsection*{One tower for different phases}
\begin{figure}[!t]
	\includegraphics[width=\columnwidth]{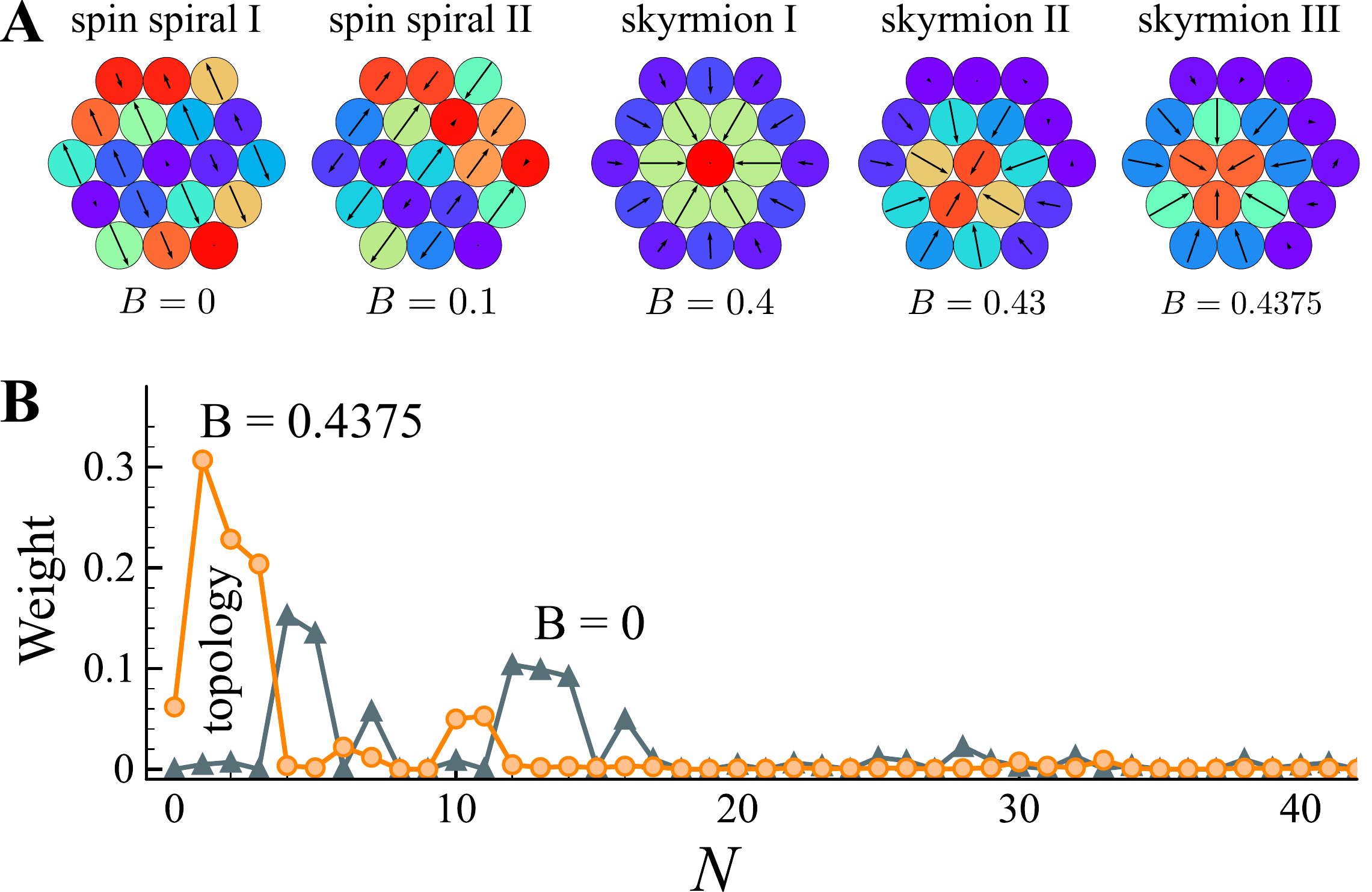}
	\caption{(A) Examples of classical spin spiral and skyrmion magnetic configurations stabilized at different magnetic fields $B$ that can be reproduced with high fidelity, ${\mathcal{F} = 0.95}$, 0.96, 0.97, 0.97, 0.97 (from left to right) by using the eigenspectrum of the quantum Hamiltonian~\eqref{Ham} with the fixed set of model parameters ${J = 0.5}$, ${|{\bf D}| = 1}$, and ${B = 0.44}$. (B) Comparison of the contributions to approximations of the spin spiral (${B=0}$) and skyrmion (${B=0.44}$) configurations.
	\label{types} }
\end{figure}

The results presented above justify the possibility to relate the eigenspectrum of the quantum spin Hamiltonian~\eqref{Ham} to the classical order obtained for the particular value of the inter-spin interaction and the magnetic field. 
Now we are in the position to demonstrate the power of the proposed approach in imitating different trivial (spin spiral) and non-trivial (skyrmion) topological structures, simultaneously. 
For this purpose we consider the same set of ${N = 105}$ eigenstates obtained for the quantum model with the parameters ${J = 0.5}$, ${|{\bf D}| = 1}$, ${B = 0.44}$. Some examples presented in Fig.~\ref{types} confirm the high fidelity of the reconstruction of completely different coherent states with the same set of quantum states.
Surprisingly, we find that coherent states corresponding to topologically trivial and non-trivial classical configurations can be obtained with the same set of eigenfunctions of the quantum spin Hamiltonian~\eqref{Ham} that reveals the quantum skyrmion state. 
The comparison of the contributions to the corresponding approximations of the classical spin spiral and skyrmion configurations is presented in the bottom panel of Fig.~\ref{types}. 
These two profiles are clearly different, and one can conclude that the topology of a classical spin texture
related to the first few (${N = 1, 2}$,~and~3 excited) levels of the quantum spectrum.    

This finding suggests a new approach to the problem of the topological protection of the classical skyrmionic structures. As it was shown previously, no continuous deformation between topologically inequivalent configurations exists in an idealized continuum consideration.
In turn, in real materials, in which magnetic skyrmions of nanosize are experimentally observed, the magnetic density is strongly localized and fully associated with concrete transition metal atoms. In this case the topologically protected configurations are separated from others by only finite energy barriers. Recent experimental results reported in Ref.~[\onlinecite{topology}] have demonstrated that the skyrmionic structures with a non-zero topological charge are characterized by longer lifetimes than trivial bubble structures, which gives evidence of the topological stability in a real discrete system. According to the results we obtained, the non-trivial topological structure of the classical magnet is fully defined by a unique combination of the eigenstates of the corresponding quantum system.

\subsection*{Tower of classical states}
The theoretical analysis presented above revolves entirely around reproducing the classical state with low-lying quantum states, and one might think that the quantum-classical connection in the Anderson's approach works only in one way. However, it is not the case. A simple justification can be demonstrated, for example, for the antiferromagnetic configuration of two spins ${S = \frac{1}{2}}$. 
On one hand, the antiferromagnetic state can be expressed through a combination of the singlet wave functions ${\Psi_{A_1}=\frac{1}{\sqrt{2}} (\Ket{\uparrow \downarrow} - \Ket{\downarrow \uparrow})}$ and ${\Psi_{A_2}=\frac{1}{\sqrt{2}} (\Ket{\uparrow \downarrow} + \Ket{\downarrow \uparrow})}$, which corresponds to the tower of quantum states. 
On the other hand, the singlet wave function entangles two coherent antiferromagnetic states ${\Psi_{T_1}=\Ket{\uparrow \downarrow}}$ and ${\Psi_{T_2}=\Ket{\downarrow \uparrow}}$. 
It means that one can construct two different types of towers: the tower of the classical states to reproduce the quantum state and, \emph{vice versa}, the tower of quantum wave functions to reconstruct the antiferromagnetic classical configuration. 
Thus, Anderson's approach supports both, classical-quantum and quantum-classical channels of characterization of the system, which fundamentally distinguishes this approach from measurement-based theories that assume complete or partial collapse of the quantum wave function upon measurement. 

\begin{figure}[!t]
	\includegraphics[width=\columnwidth]{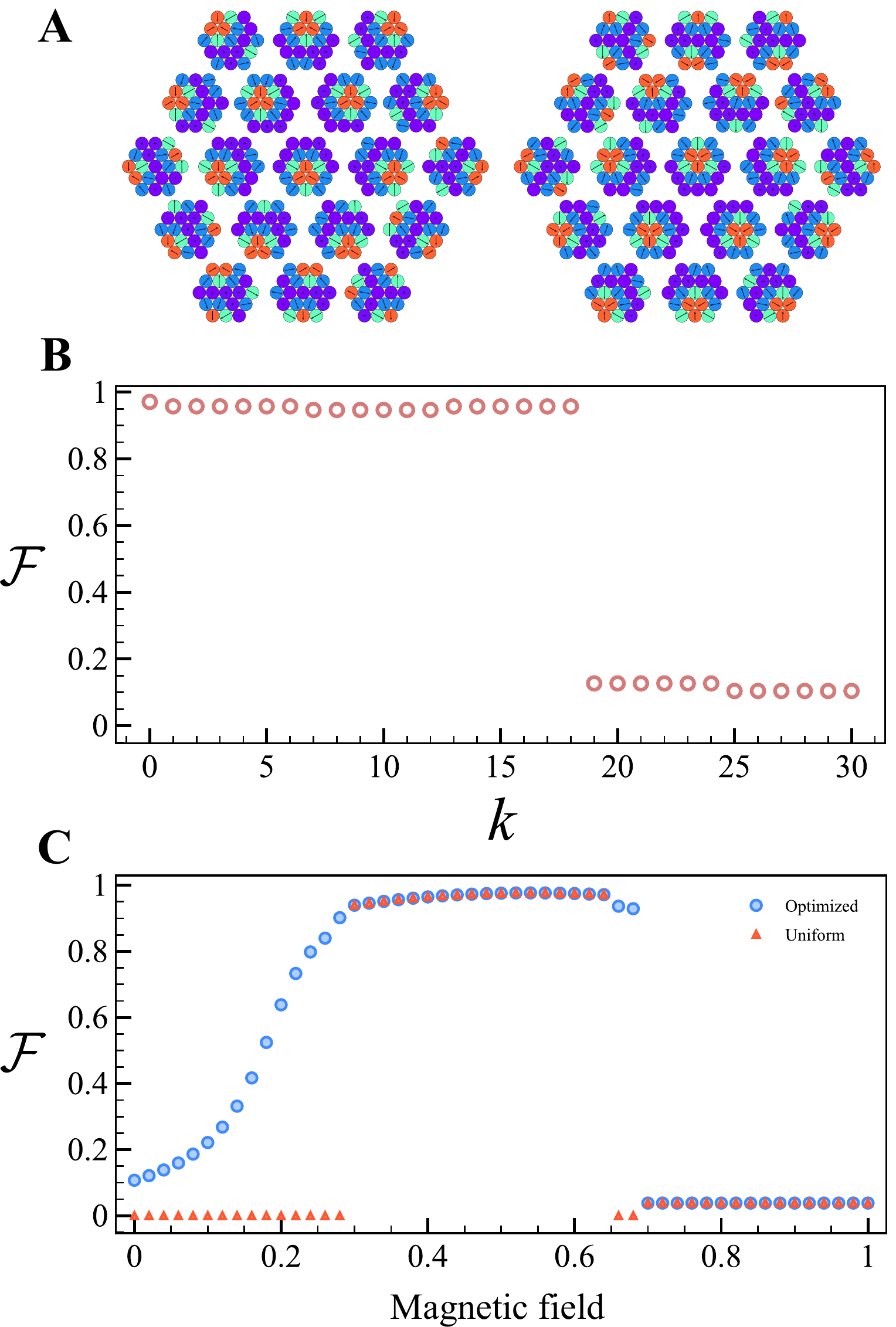}
	\caption{\label{classic_towers} (A) 38 classical skyrmionic configurations obtained at B =0.44 and used to define the tower of classical states, Eq.\ref{quantum_class}. (B) Reconstruction fidelities of the ground and excited eigenfunctions of the quantum Hamiltonian, Eq.\ref{Ham} calculated at B = 0.44. The reconstructions are performed  with optimization of the tower of classical states, Eq.\ref{quantum_class}. (C) Calculated fidelities (circles) describing the quality of optimizing the quantum ground states at different magnetic fields with the tower of classical states. Triangles correspond to the approximation of Eq.\ref{quantum_class} taken with $a_1 = a_2 =..= a_{38} = 0.1078068$.}
\end{figure}

To explore the possibility of reconstructing the quantum states, which are the eigenfunctions of the quantum spin Hamiltonian, with classical solutions we built a tower of coherent states on the basis of the classical skyrmionic solution obtained at ${B = 0.44}$.  
This magnetic skyrmion is visualized in Fig.~\ref{types} A (skyrmion III), and its properties regarding the spin-spin correlation functions were discussed in a previous work~\cite{Quantum_skyrmion}. 
Initially, the tower of states is built up as a random superposition of the coherent states corresponding to the considered classical skyrmion and its replicas obtained applying translational and rotational symmetry operations as visualized in Fig.~\ref{classic_towers}\,A. 
The number of coherent states contributing to the tower depends on the specific classical configuration and in our case is equal to 38:  
\begin{eqnarray}
\label{quantum_class}
	|\Psi^{\rm cl}_{\rm A} \rangle =
		a_1
		\raisebox{-0.13cm}{
			\includegraphics[width=0.7cm]{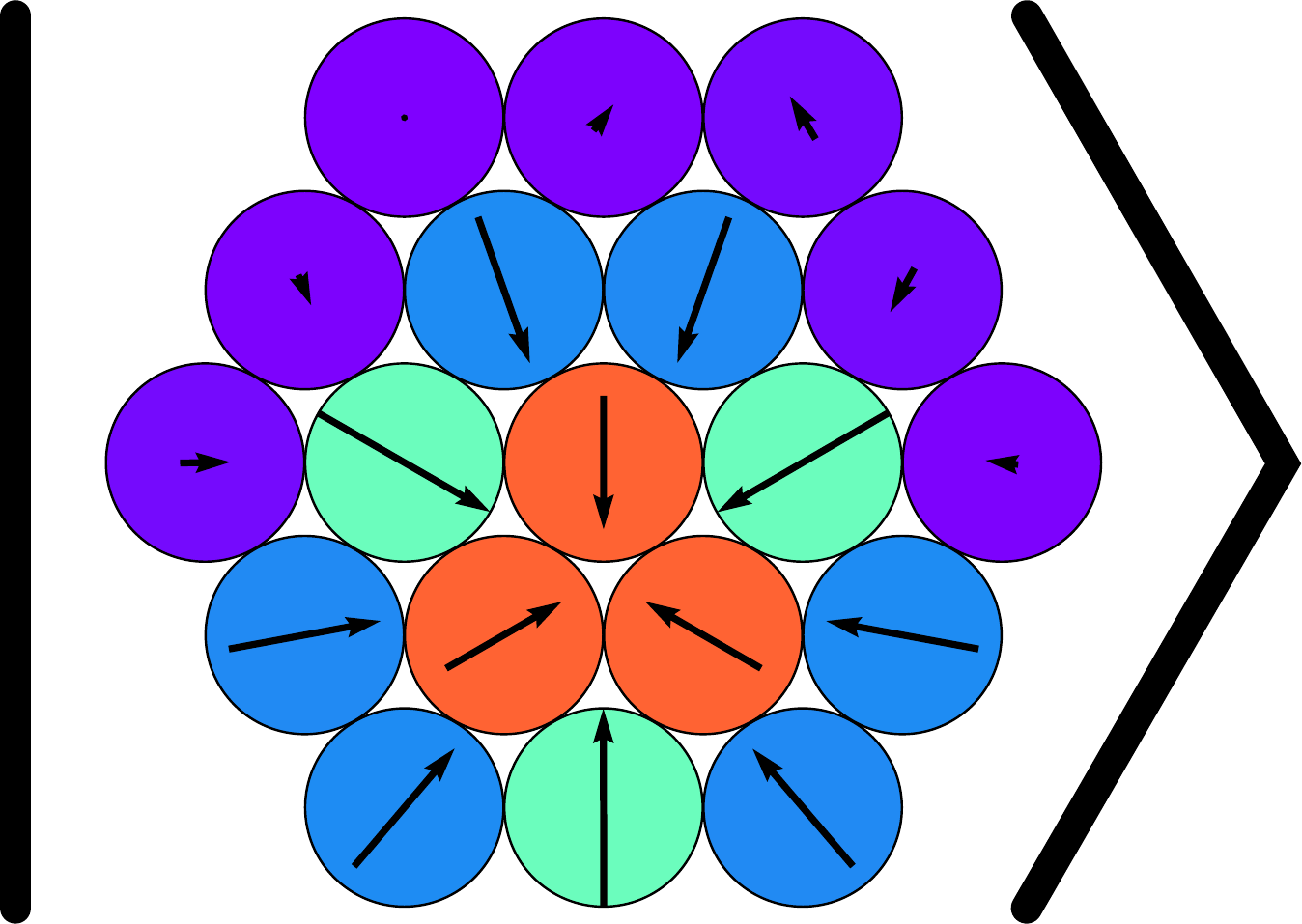}
		}  \negmedspace +
		a_2	\raisebox{-0.13cm}{
			\includegraphics[width=0.7cm]{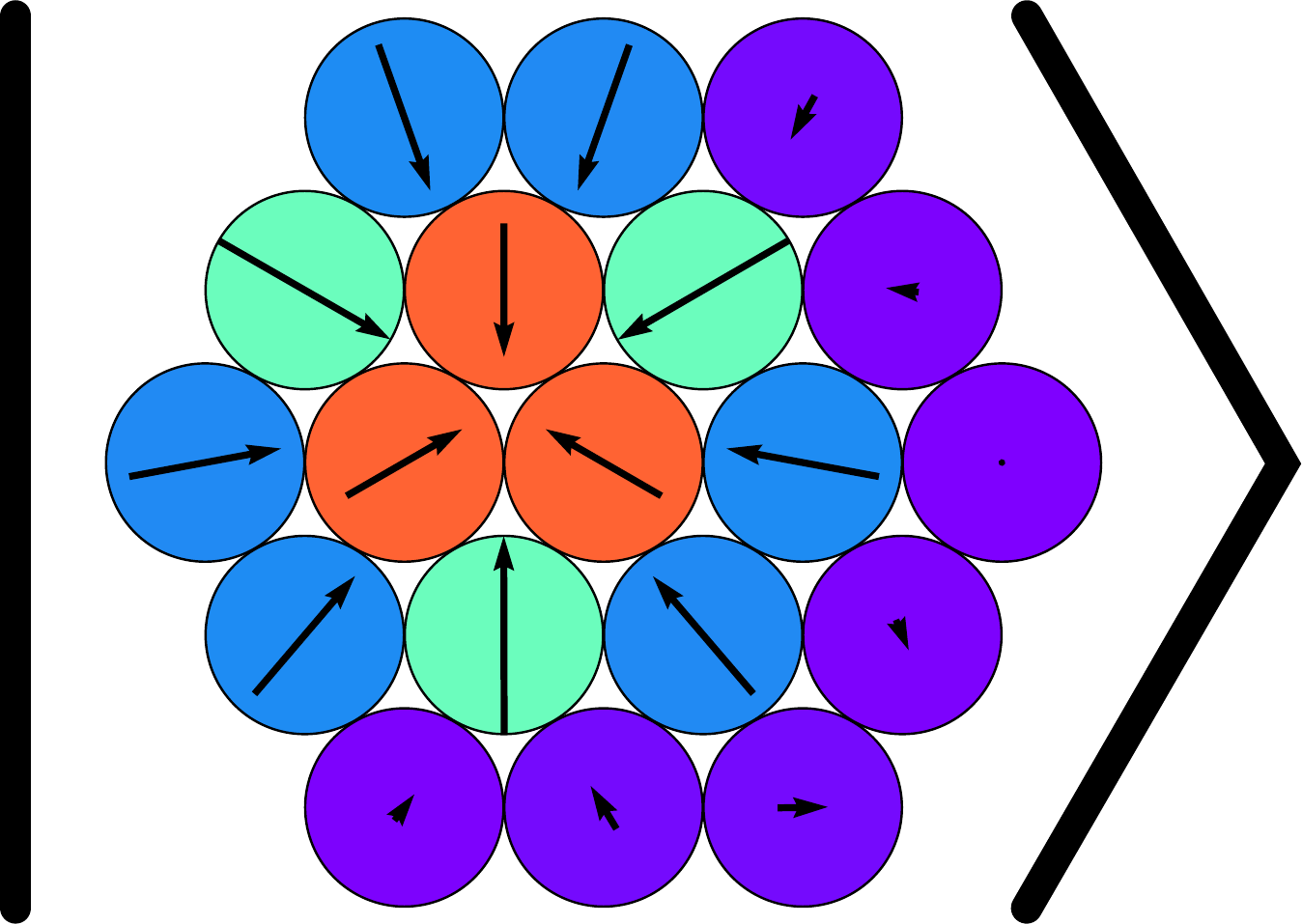}
		} \negmedspace + \ldots + 
		a_{38}	\raisebox{-0.13cm}{
			\includegraphics[width=0.7cm]{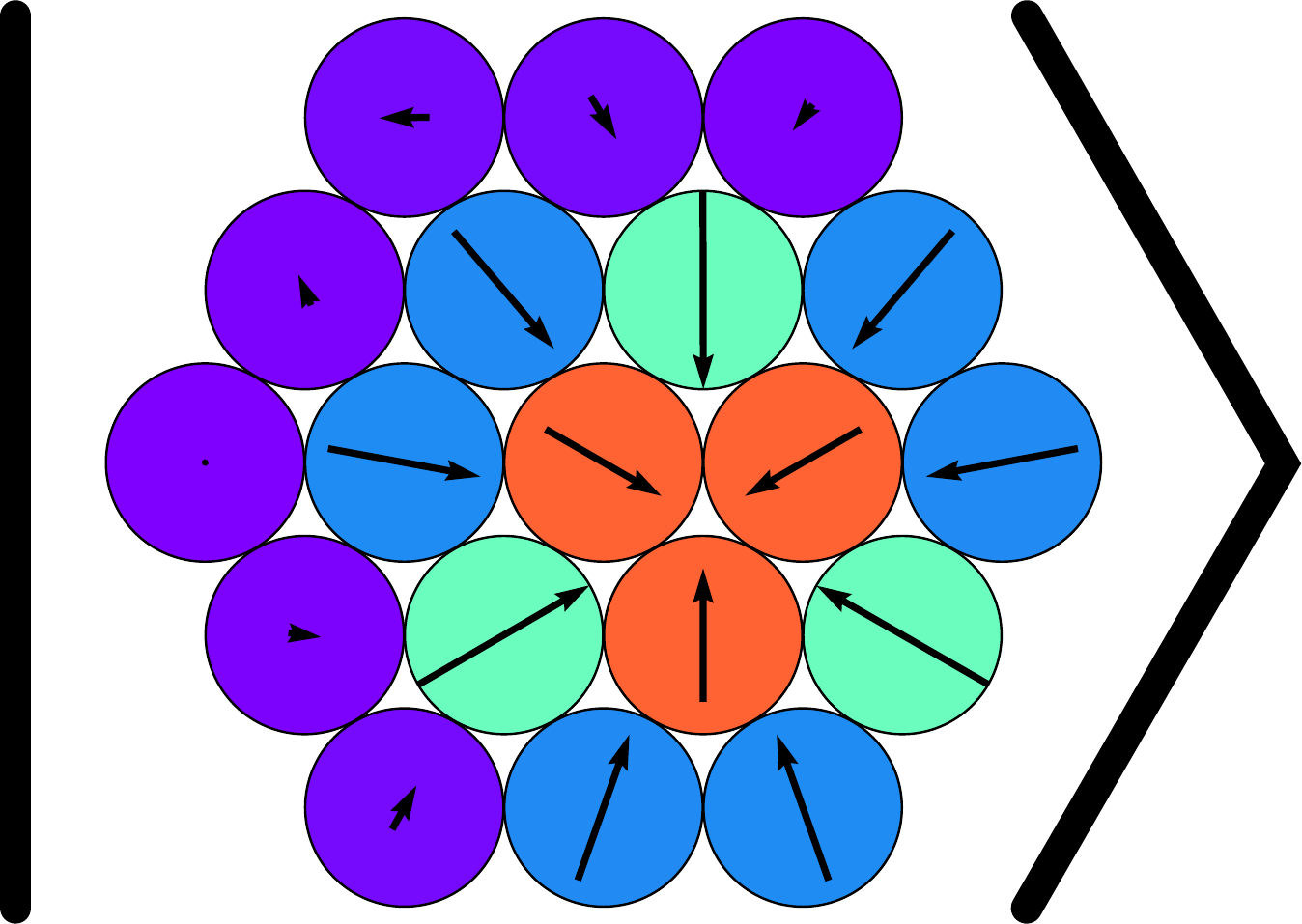}
		},
\end{eqnarray}
where $a_i$ is the amplitude of the $i$-th coherent state. 
One should note that $\sum_{i}a_i^2 \ne 1$, since coherent states in Eq.~\ref{quantum_class} are not orthogonal to each other. Nevertheless, the state $\Psi^{\rm cl}_{\rm A}$ remains to be normalized to 1 in the $\sigma^z$ basis.

The constructed tower is optimized with the gradient-descent approach to get maximum possible overlap with one of the states from the low-lying part of the eigenspectrum. 
For demonstration purpose, we chose the eigenspectrum of the quantum spin Hamiltonian~\eqref{Ham} calculated for the same value of the magnetic field ${B=0.44}$. 
From Fig.~\ref{classic_towers}\,B one can see that the constructed classical tower reveals high fidelity (more than 95\%) in reconstructing the eigenstates with ${k=0,\ldots,18}$ that belong to the low-energy part of the eigenspectrum below the energy gap.
In contrast to the low-energy eigenfunctions, the calculated fidelities for the excited states above the gap are nearly zero.

The optimization of the quantum ground states obtained at different values of magnetic field (circles in Fig.~\ref{classic_towers}\,C) reveal high fidelities close to 1 within the skyrmionic phase for ${0.3 \leq B < 0.66}$~\cite{Quantum_skyrmion}. Interestingly, for ${B < 0.3}$ one can find a combination of coefficients in Eq.~\eqref{quantum_class} with gradient-descent method such that there is a non-zero overlap between the quantum ground state and the constructed tower of classical skyrmionic configurations. 
Even at ${B = 0}$ the fidelity is not identically zero and has the value of about 0.1, which is a direct evidence that there is a quantum superposition of the spin spiral and skyrmion states in the range of fields ${0 \leq B < 0.3}$. 

Another important result is that the optimized coefficients within the skyrmionic phase (${0.3 \leq B < 0.66}$) are practically the same for each term in Eq.~\eqref{quantum_class}. Triangle symbols in Fig.~\ref{classic_towers}\,C denote the overlap between the quantum ground state calculated for a specific magnetic field and the wave function $\Psi^{\rm cl}_{\rm A}$ taken with ${a = a_1 = a_2 =..= a_{38} = 0.1078068}$. 
One can see that using such a uniform state
\begin{eqnarray}
\label{quantum_class1}
	|\Psi^{\rm cl}_{\rm U} \rangle =
		a (
		\raisebox{-0.13cm}{
			\includegraphics[width=0.7cm]{basis_state_1}
		}  \negmedspace +
			\raisebox{-0.13cm}{
			\includegraphics[width=0.7cm]{basis_state_2}
		} \negmedspace + \ldots + 
			\raisebox{-0.13cm}{
			\includegraphics[width=0.7cm]{basis_state_3}
		}),
\end{eqnarray}
we can approximate any ground state within the skyrmionic phase with high fidelity. 
In other words, it means that the quantum ground state within the skyrmionic phase is almost insensitive to the external magnetic field. This specific choice of coefficients gives the wave function for a pure skyrmion state, so fidelity of this pure skyrmion state with the state that is a superposition of spin spirals and skyrmions is zero as we see for $B<0.3$.

Thus, a highly-entangled wave function $|\Psi_{0} \rangle$, the ground state of the quantum spin Hamiltonian~\eqref{Ham}, can be expressed as a simple superposition of trivial coherent states that correspond to the same classical magnetic configuration and differ from each other only by translation or rotation operations. 
We believe that this result is remarkable. 
First of all, it directly demonstrates the connection between quantum and classical skyrmion states. Previously, the relation between quantum and classical skyrmionic solutions was established only on the level of observables, such as the magnetization, the spin structural factors, and the scalar chirality. 
Further, Eq.~\eqref{quantum_class} suggests a distinct way for simulating and storing large-scale quantum topological states. 
They can be efficiently prepared on the basis of the classical solution thus avoiding hard-to-do exact diagonalization procedure, for which the current limit is a quantum system of 50 spins~\cite{Lauchli}.  

It is also worth discussing the connection of the quantum state decomposition onto coherent states realized within Eq.~\eqref{quantum_class} and the results of projective measurements of $\Psi_{0}$ reported in our previous work~[\onlinecite{Quantum_skyrmion}]. 
Looking at Eq.~\eqref{quantum_class}, one might think that one of the classical replicas of a skyrmion should be observed in the experiments after the measurement. 
However, it is not the case. 
A single projective measurement in $\sigma^z$ basis results in a basis function that can be attributed to one or more coherent states~\eqref{target} simultaneously. 
That is why in Ref.~[\onlinecite{Quantum_skyrmion}] when measuring $\Psi_0$ a sequence of basis states that are fully unstructured with respect to the classical skyrmion profile was observed. 
Averaging over such measurements leads to a uniform magnetization. 
To amplify the contribution of one particular coherent state it is necessary to perform the measurements by using the corresponding local bases for each spin in the system. In principle, such a procedure can be realized in quantum computing. In this case one will get the basis state $|000..0 \rangle$ or $|\uparrow \uparrow \uparrow .. \uparrow \rangle$ with the probability $\sim 0.01$, which can be considered as a finite value in comparison with practically zero contribution of that basis function in the global $\sigma^z$ basis.

Below we will discuss a purely quantum mechanism that allows to preselect a specific coherent state from Eq.~\eqref{quantum_class} before the measurements. 

\subsection*{Comparison to the case of quantum antiferromagnets}
Previous studies~\cite{Misguich, Frederic, Bernu} aiming at detecting broken-symmetries in quantum antiferromagnets with Anderson towers are mainly based on the group-theoretical calculations that only allow one to identify the contributing eigenstates by their symmetries. However, a combination of these eigenstates taken with some amplitudes into a concrete coherent wave function and exploring its properties by calculating different observables are beyond the capabilities of the group-theoretical approach. Our protocol for constructing Anderson towers allows for the direct numerical estimation of the contribution of individual eigenstates to broken-symmetry wave functions, which, to the best of our knowledge, was not done up to date for quantum antiferromagnets. In this respect we would like to cite the results of the paper~\cite{Frederic_tower} in which an analytic expression for the tower decomposition coefficients in the specific limit of the Bose-Hubbard model parameters was derived. Based on this the authors have shown that one can find a signature of a nematic order when constructing the TOS for the quantum system being in the superfluid ground state of spin-1 bosons. Thus, it is instructive to implement our approach to explore canonical quantum antiferromagnets and compare them to the case of the quantum skyrmion. 

\begin{figure}[!t]
	\includegraphics[width=0.9\linewidth]{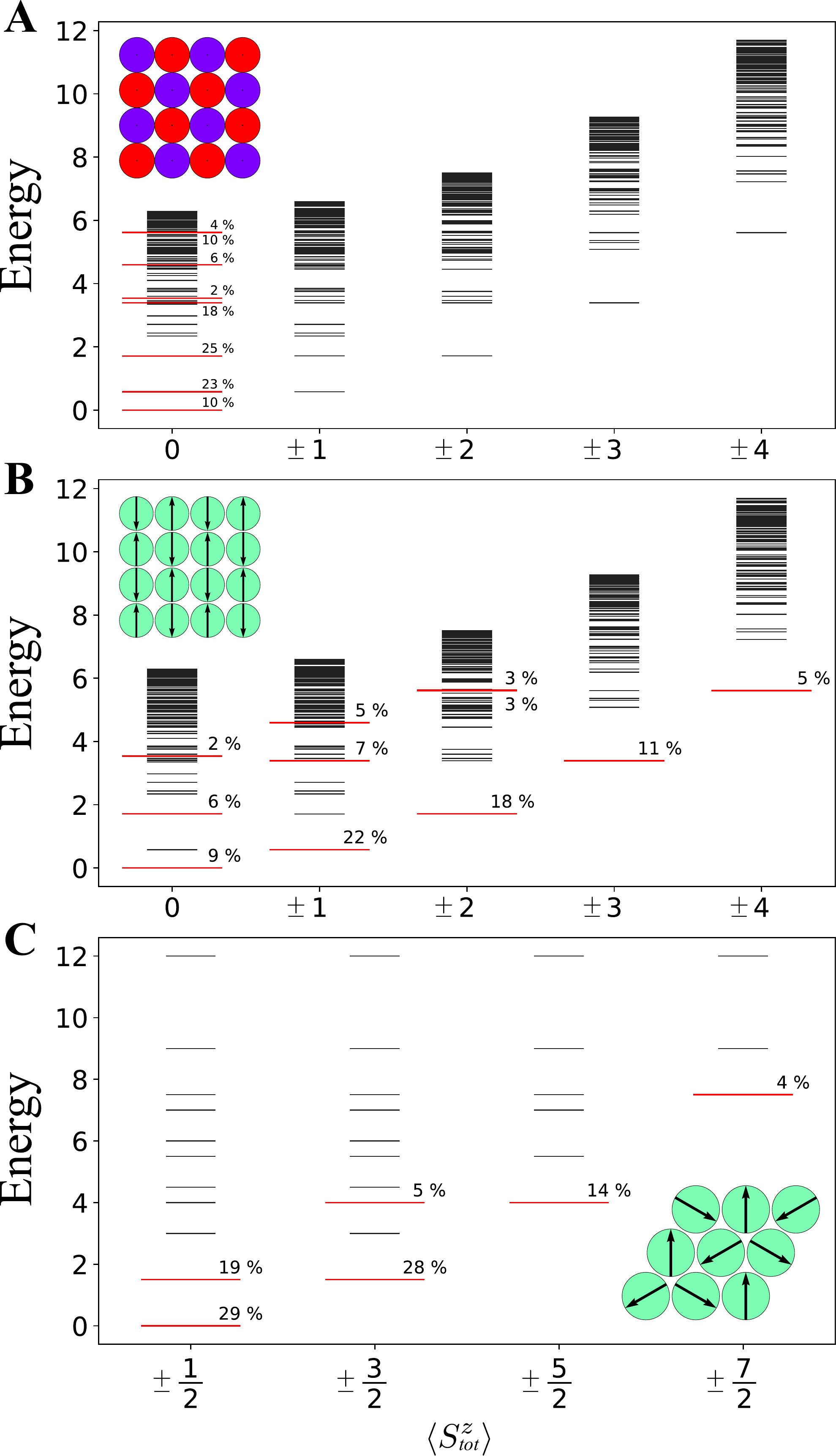}
	\caption{\label{anti} Eigenspectra (black lines) and largest contributions to TOS (red lines) obtained for $z$-oriented AFM (A), $y$-oriented AFM (B) and 120$^{\circ}$ AFM (C) orderings. The simulations were performed with 4x4 square (A and B) and 3$\times$3 triangular (C) supercells. For the sake of the visualization the eigenstates giving 1\% or less contribution to the TOS are not highlighted in red.}
\end{figure}

Fig.~\ref{anti} visualises the contribution of the eigenstates to the Anderson towers for quantum antiferromagnets  with only isotropic exchange interaction between nearest neighbours
\begin{align}
\hat {\rm H}_{\rm AFM} = \sum_{ij} J_{ij} \hat{\bf S}_i  \cdot  \hat{\bf S}_j, 
\label{antiHam}
\end{align}
defined on the square and triangular supercells with periodic boundary conditions. 
First, let us discuss the results obtained for the case of the square lattice antiferromagnet. 
If the directions of the magnetic moments in the classical texture are collinear to the $z$ axis, all the non-zero contributions to the Anderson tower come from the states with zero total spin (Fig.~\ref{anti}\,A). A non-monotonous behaviour as the energy of the state increases represents another peculiarity of the constructed tower. While the largest weights $|\alpha_n|^2$ are provided by the excited states belonging to the first and second eigenlevels, the ground state gives about 20\% of the coherent state. If the classical antiferromagnetic structure is along the $y$ axis, as shown in Fig.~\ref{anti}\,B, the distribution of the leading contributions over spin sectors becomes completely delocalized in comparison with the $z$ axis case. Such a tower is mainly built by the lowest energy states in each spin sector. Basically, the obtained states structure agrees with the conventional picture of TOS~\cite{Wietek} for the Heisenberg model on a square lattice. However, we additionally observe significant contributions from the excited states. Such a deviation from the standard TOS scenario may be explained by the small size of the simulated systems.    

To examine whether these contributions of the excited states to the TOS are a unique property of the square lattice antiferromagnet, or they can also be revealed in Heisenberg systems with different scheme of isotropic interactions, we consider the spin model~\eqref{antiHam} on a triangular lattice, for which the target coherent wave function corresponding to the classical order is assumed to be a 120$^{\circ}$ antiferromagnetic state~\cite{Villain}. According to calculations on a 36-site cluster, the ground state of the quantum Heisenberg Hamiltonian subjected to frustration on the triangular lattice has a very small order, if any, as follows from the decay of the spin-spin correlation function at large inter-spin distances~\cite{Leung}. This frustration requires simulating quantum systems of large enough sizes in order to perform an extrapolation of the system's properties to the thermodynamics limit. 
Advanced techniques such as neural quantum state~\cite{NQS} potentially allow for modelling large quantum systems. 
However, they do not provide the required accuracy in determining the ground and the excited eigenstate energies of frustrated magnets~\cite{Bagrov}, which makes it inevitable to use exact diagonalization methods. 

Such fundamental limitations on the size of the frustrated systems encouraged an active use of the TOS concept to explore the presence of magnetic order in the triangular lattice Heisenberg antiferromagnet~\cite{Bernu, Nishimori, Azaria}.  
In particular, as shown in Ref.~[\onlinecite{Bernu}], there is a whole set of excited states which can constitute the tower of state. Our numerical results presented in Fig.~\ref{anti}\,C support this symmetry-based conclusion. 
For the in-plane 120$^{\circ}$ N\'eel ordering simulated with a 3$\times$3 supercell the eigenspectra resolved in total spin numbers is symmetrical. The largest contributions of about 29\% of the target coherent wave function come from the ground eigenstates in spin sectors with ${S^z_{tot} = \pm \frac{1}{2}, \pm \frac{3}{2}}$. It is clearly seen from Fig.~\ref{anti}\,C that the excited states of the same sectors provide non-negligible contributions from 5 to 19\%.  

Thus, one can define the eigenfunctions candidates that build the TOS with group-theoretical consideration or by using the gradient-descent approach we proposed in this work. 
Importantly, our analysis of the quantum skyrmions (Fig.~\ref{weight}) and quantum antiferromagnets (Fig.~\ref{anti}) has demonstrated that the coherent state representing the particular classical configuration can be reconstructed with only specific combination of the eigenstates.
However, in the thermodynamics limit, following the original idea by Anderson, these states should collapse to the highly degenerate ground state manifold, which means that the quantum system in question can be described by an arbitrary superposition of the considered wave functions.
Thus, one faces a critical problem that was not addressed in the previous works on TOS, namely the mechanism responsible for the transformation of the random composition of the TOS eigenfunctions onto their specific combination. In the next section we will take a step forward and propose a concrete scenario describing such a transformation using the developed gradient-descent protocol.

\subsection*{Loss function for decoherence} 
\begin{figure}[t]
	\includegraphics[width=\columnwidth]{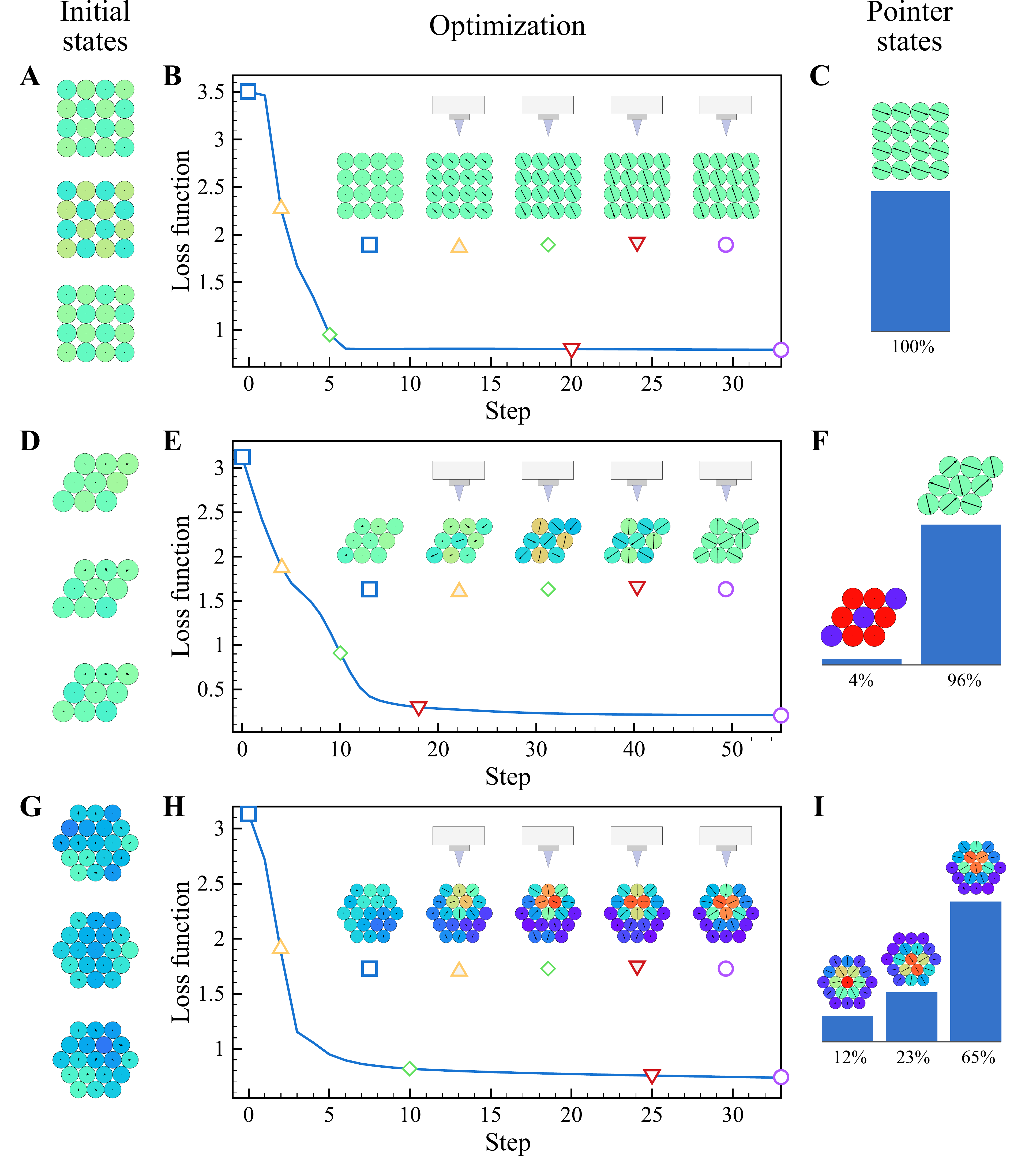}
	\caption{\label{decoherence} Searching for pointer states. Examples of magnetic structures corresponding to the different realizations of the random state $\Psi_{\rm R}$ generated with TOS eigenfunctions in the case of square-lattice (A) and triangular-lattice (D) Heisenberg models as well as for quantum skyrmion (G). (B, E, H) Loss functions describing the decrease of the entanglement in the system in question due to decoherence. The insets show the evolution of the magnetic structures during the optimization. (C. F. I) Examples of pointer states - classical configurations that are preselected by the decoherence and revealed in the quantum state measurements. Histograms denote the percentage of realization of a certain pointer state in 100 independent numerical experiments for each spin model.}
\end{figure}

The tower of states analysis is a very important approach to detect the spectral structure for several scenarios of symmetry breaking in the thermodynamics limit, but it does not unveil all the details of such a transformation. 
To demonstrate this we have prepared 100 trial wave functions $\Psi_{\rm R}$ representing a random superposition of the TOS eigenstates (the corresponding eigenlevels are denoted with red in Fig.~\ref{anti}) for both, square and triangular lattice quantum antiferromagnets. In other words, this random wave function is given by the following expression:
\begin{eqnarray}
\Psi_{\rm R} (\mbox{\boldmath$\alpha$}) = \sum_{n=0}^{k} \alpha_n \Psi^{\rm TOS}_{n},
\end{eqnarray}
where $\alpha_n$ are random complex coefficients. One can think of constructing such random superpositions as a finite-size supercell imitation of the thermodynamics limit in which these TOS eigenstates should form a degenerate ground state manifold, which means that the quantum system can be found in an arbitrary combination of such eigenfunctions. 
For the square lattice quantum antiferromagnet we consider the case where all the TOS levels are concentrated in ${S^z_{tot} = 0}$ sector (Fig.~\ref{anti}\,A). 
The eigenstates used to construct $\Psi_{\rm R}$ for the triangular lattice Heisenberg model are collected from different spin sectors (Fig.~\ref{anti}\,C). 
Magnetization textures obtained as a result of projective measurements of such random superpositions ($\Psi_{\rm R}$) are presented in Figs.~\ref{decoherence}\,A~and~D. 
Triangular plaquette Heisenberg antiferromagnets are characterized by non-uniform magnetization profiles $\langle \hat{S}^z_{i} \rangle$ (Fig.~\ref{decoherence}\,A) with small random $\langle \hat{S}^x_{i} \rangle$ and $\langle \hat{S}^y_{i} \rangle$ components fluctuating from sample to sample. 
In turn, one can recognize an antiferromagnetic pattern in the case of the square lattice.
However, the observed lengths of the magnetic moments are negligibly small compared to what one would expect for a classical configuration. These results raise an important question about the connection between $\Psi_{\rm R}$ and the coherent antiferromagnetic states $\Psi_{\textrm {N\'eel}}$, which was not discussed in the previous works concerning the TOS, but at the same time it would be a demonstration of the internal consistency of the whole TOS theory no more no less.  

Since the main focus in our work is on the reconstruction of the classical magnetic structures, it is important to discuss details of building random skyrmionic TOS and compare them with coherent skyrmion wave function. As shown above, the eigenspectrum of the skyrmionic Hamiltonian~\eqref{Ham} is characterized by 19 low-lying eigenstates that compose 4 eigenlevels and are well separated by the sizable energy gap from the rest of the spectrum. 
Exact diagonalization calculations (see the Methods section) show that the energy gap varies from 0.2 to 0.7 in units of DMI within the skyrmionic phase. 
Comparing this to the energy width of the set of 19 low-lying states that is about 0.03 in units of DMI, one can consider these eigenstates to be nearly degenerate. The low-temperature physics of the quantum skyrmion system can be described with a random superposition of only the 19 low-lying eigenstates. Following the antiferromagnets consideration we have generated 100 different such superpositions and measured their scalar chirality. Remarkably, in all the cases, the total chirality~\eqref{QPsi} calculated for $\Psi_{\rm R}$ has the value of 0.51, which was also found for the ground state of the same Hamiltonian~\cite{Quantum_skyrmion}. At the same time, the generated ensembles of $\Psi_{\rm R}$ in the case of skyrmions are characterized by featureless magnetization textures (Fig.~\ref{decoherence}\,G) fluctuating from sample to sample, which is similar to antiferromagnets. It contradict the results of real-space magnetization imaging experiments, such as spin-polarized scanning tunneling microscopy~\cite{Wiesendanger, Wiesendanger1} and Lorentz transmission electron microscopy~\cite{Yu} experiments in which one routinely observes reproducible skyrmionic vortex-like profiles. 

Thus, we arrive at the famous problem of the transition from the quantum realm characterized by a superposition of alternatives (outcomes) to the classical reality with a single outcome, which haunts researchers since the very foundation of quantum mechanics~\cite{Bohr, Everett, Wheeler}. However, the quantum-classical transition we deal with in this work is a specific one.  
The performed numerical experiments on measurements of the magnetization density in skyrmionic systems and antiferromagnets unequivocally testify that there should be an additional transformation of the system's wave function before the measurement, which would give a classical spin configuration as an outcome of the measurement. 
Now, we have to establish the relation of the TOS approach to the decoherence program~\cite{Joos2,Joos1,Zurek0,Zurek1,Zurek3,Zurek2}. 
First of all, according to the decoherence theory, macroscopic systems are never isolated from their environments, which automatically makes the TOS approach, which predicts the properties of an isolated quantum system in the thermodynamic limit, incomplete without taking into account the influence of the environment. The decoherence leads to the environment-induced superselection of states (pointer states), which remain stable in the presence of the environment. The environment has a little effect on the pointer states, since they are already classical. Thus, following this definition, we would like to stress that the coherent (target) wave functions corresponding to the classical skyrmionic or antiferromagnetic textures we reconstruct within the TOS framework can be associated to such pointer states introduced in the decoherence theory. For simpler model situations, the transition from quantum singlet state to classical N\'eel state under the effect of the environment was studied in Refs.~[\onlinecite{singlet_to_neel1}]~and~[\onlinecite{singlet_to_neel2}].  

In general, the search for the pointer states can be realized within the following scheme. First, one has to define a general density matrix $\rho_{\mathcal{SE}}$ that describes an entangled superposition of the quantum system in question ($\mathcal{S}$) and the environment ($\mathcal{E}$). Then, it is necessary to compute the reduced density matrix for the quantum system by tracing the environmental part out: ${\rho_{\mathcal{S}} = {\rm Tr}_{\mathcal{E}} \rho_{\mathcal{SE}}}$. 
All possible states of the system will be involved in $\rho_{\mathcal{S}}$ defined as a function of time.  
Thus, minimizing von Neumann entropy for $\rho_{\mathcal{S}}(t)$ allows one to detect the pointer states. However, having introduced an environment one should make some assumption on its properties, which leads to the loss of generality of the performed analysis. Besides, from the numerical point of view a simulation of the environment to detect the pointer states looks somehow unrealistic since it should be characterized by an exponentially larger number of degrees of freedom than those describing the investigated quantum system itself.

Fortunately, the wave functions we consider within the TOS theory to describe quantum antiferromagnets and skyrmions contain pointer states by construction. Thus, the influence of the environment on the quantum system can be considered as an optimization of $\Psi_{\rm R}$ as a function of \mbox{\boldmath$\alpha$} towards a coherent (pointer) state. Importantly, the environment could be taken into account on the level of a loss function without explicitly introducing it into the model. The discussion above on the decoherence theory makes the choice of the loss function obvious:
 \begin{eqnarray}
 \label{decoh}
\mathbb{E}_{\rm decoh} (\mbox{\boldmath$\alpha$}) = -{\rm Tr} \rho_{\rm A} (\mbox{\boldmath$\alpha$}) \log_2 \rho_{\rm A} (\mbox{\boldmath$\alpha$}),
\end{eqnarray}
where ${\rho_{\rm A} (\mbox{\boldmath$\alpha$}) = {\rm Tr}_{\rm B} \rho_{\rm AB} (\mbox{\boldmath$\alpha$})}$ is the reduced density matrix for a half-system biparition into parts A and B. This loss function is nothing but von Neumann entanglement entropy.  

The use of such loss function for optimizing $\Psi_{\rm R}$ allows us to describe the decrease of the uncertainty due to the contact of the quantum state with the environment. Some examples of the evolution of random states $\Psi_{\rm R}$ describing quantum antiferromagnets and skyrmions within gradient-descent procedure are presented in Figs.~\ref{decoherence}\,B,\,E~and~H. Depending on the initial random combination of the TOS eigenfunctions different alternatives of the classical outcome can be realized. In the case of the square lattice Heisenberg antiferromagnet the symmetry of the TOS eigenstates allows for the formation only one type of pointer state that is the 180$^{\circ}$ N\'eel configuration (Fig.~\ref{decoherence}\,C). At the same time there are two alternatives allowed by the symmetry of triangular lattice Heisenberg model, namely 120$^{\circ}$ N\'eel and collinear stripe phases (Fig.~\ref{decoherence}\,F). The latter is characterized by the larger energy than the former one for the considered spin Hamiltonian~\eqref{antiHam}.
However, the accounting for the next-nearest-neighbour exchange interactions in the model Hamiltonian is known to change the balance between these classical configurations~\cite{Pierre}. 

 \begin{figure}[t]
	\includegraphics[width=0.8\columnwidth]{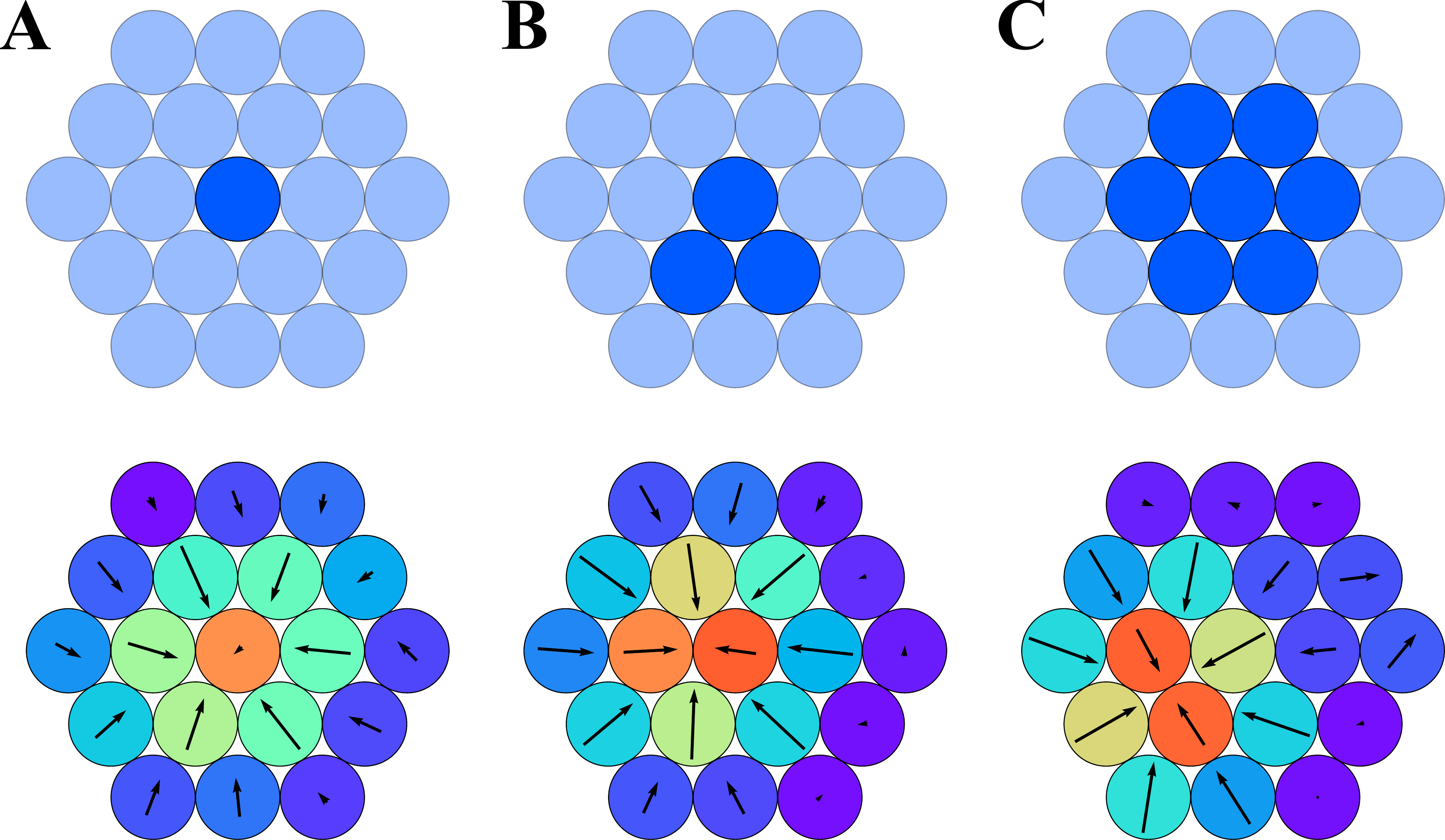}
	\caption{\label{vonNeumann} Comparison of the optimization results obtained with different number of spins in the subsystem A, Eq.\ref{decoh} used to extract information on entanglement of the system in question. In the upper panel, the spin clusters used for the definition of the reduced density matrix $\rho_{A}$ in the optimization procedure Eq.\ref{decoh} are highlighted. The bottom panel gives  examples of the resulting skyrmionic structures induced by the minimization of the entanglement entropy defined for the corresponding clusters.}
\end{figure}

The period of the antiferromagnetic structures is shorter than that of the skyrmion one.
The square 4$\times$4 and triangular 3$\times$3 supercells we used can be safely divided into smaller fragments (clusters) locally reproducing the corresponding antiferromagnetic configuration between nearest spins. It means that the reduced density matrix is defined on the system's fragment that is expected to contain complete information on the magnetic configuration stabilized in the classical case. For the quantum skyrmion system the situation is more complicated. 19-site plaquette can host only one skyrmion~\cite{Quantum_skyrmion}. It means that the subsystem A of 9 sites for which we calculate $\rho_{\rm A}$ covers roughly a half of the skyrmion in the classical case. Judging by the fact that the optimization of entanglement entropy calculated for the local fragment of the quantum system restores the classical skyrmionic structure, one can conclude that the corresponding reduced density matrix compresses a complete information on this three-dimensional topological texture.

To elaborate on this point and to probe the limit of such quantum compression of the classical information we performed the quantum-classical optimization described in Eq.~\eqref{decoh} by choosing subsystems of different shapes and different number of spins involved into reduced density matrix. It follows from Fig.~\ref{vonNeumann} that the symmetry and size of the subsystem used to probe the entanglement strongly affect the quality of the resulting classical state. Nevertheless, even the single-site entropy contains enough information to get recognisable skyrmionic pattern, albeit with some symmetry distortion.  

The survival of the most robust, that is, pointer states in an environment was called by Zurek ``quantum Darwinism''~\cite{Zurek2}. It is worth to mention that the word ``Darwinism'' when discussing optimization problem means probably much more than just a bright analogy. Formal correspondence between Darwinian evolution, optimization problem in machine learning, and statistical mechanics was recently discussed in Refs.~[\onlinecite{Vanchurin1}]~and~[\onlinecite{Vanchurin2}], the loss function corresponds to (minus) the logarithm of fitness in evolutionary biology and to the free energy in statistics. 
  
\section*{Discussions}
The catalog of the topologically-protected classical magnetic quasiparticles found in real or model systems is already quite extensive and is constantly supplemented. It includes different types of two-~\cite{Si111,bimeron} and three-dimensional skyrmions~\cite{Blugel1, Braun}, and related spin textures~\cite{Gobel}. One could expect that in the quantum case the list of skyrmionic structures would be much richer and more diverse. It simply follows from the fact that all the classical skyrmionic species we already know could be potentially reproduced with quantum mechanics and then on this basis various entangled superposition of different coherent skyrmionic wave functions can be created. For instance, one can formally initialize a superposition of Bloch and N\'eel skyrmions, which is impossible in the classical case. Practical realization of such manipulations with topologically-protected spin textures are in demand for creating novel quantum technologies~\cite{skyrmion_as_qubit} and ultimately requires to establish a direct conformity between the state of a quantum system and topologically-protected classical order. 

In our work, we have revealed and explored such a way to connect quantum and classical skyrmion systems by means of the combination of the Anderson tower of states and decoherence theory. It is unambiguously shown that classical non-collinear configuration of spins with non-zero chirality can be reconstructed with the set of a few low-lying excited eigenstates of the quantum Hamiltonian containing isotropic and anisotropic exchange interactions. Importantly, the same quantum model can be used for imitating topological magnetization patterns stabilized with various microscopic mechanisms including frustration of isotropic exchange interactions~\cite{Okubo}, high-order spin couplings~\cite{Wiesendanger} and others, which proves the versatility of the proposed approach.  Having performed a cross analysis of classical and quantum towers we have shown how highly entangled quantum skyrmion wave function can be represented as a superposition of trivial coherent states corresponding to the concrete skyrmion configuration of the classical spins. In turn, the imitation of the decoherence due to the contact with environment by using a 
gradient-descent approach uncovers the evolution of the constructed tower of quantum states and reveals the transition of the highly-entangled quantum skyrmion to a set of pointer states, trivial wave functions with zero entanglement entropy corresponding to the particular classical skyrmions. This defines the mechanism through which one can observe a classical skyrmion in 
a system which is a priori quantum.      

Among the research directions that can be initiated by using our results we would like to discuss those related to machine learning that plays a crucial role in solving different real-world problems such as classification, recognition, translation, optimization and others.  The gradient-descent method we used for constructing Anderson's tower of states is one of the main workhorses in machine learning. Constructing Anderson towers of states is similar to learning in a sense that a tower of entangled quantum states is optimized to reproduce different coherent wave functions with fidelity as high as possible. And the results of such a learning can be used for the phase classification problem in the case of non-collinear topological structures. More specifically, as follows from Fig.~\ref{types}  one can distinguish the spin spiral and skyrmion phases by the decomposition of the weight profiles of the towers. In general, performing phase classification in the case of non-collinear magnets is supposed to be a hard problem that can be solved by inventing order parameters or other means including machine learning~\cite{supervised, profile, complexity, dissimilarity}.  Our results suggest a distinct way for phase classification through Anderson towers.

In the proposed realization of the Anderson idea on tower of states one can find a motif of the generative machine learning~\cite{generative}, one of the goals of which is to generate new reliable content (images, videos).  Instead of explicitly specifying a target state, we can describe its required properties using an appropriate loss function. Importantly, such properties can be related not only to some physical observables  (magnetization and spin-spin correlation functions), but also to pure theoretical physical quantities such as Shannon entropy. It opens the way to constructing different classical states as well as various quantum-classical hybrid wave functions characterized by non-zero entanglement and specific magnetization patterns. The demonstrated reconstruction of  pointer state from the tower of states by just minimising the entanglement entropy in that subspace could have interesting applications in identifying the order in systems where one knows that there is a phase transition but has no clue as to what the order could be (a problem known as hidden order): from the low energy states, one could try to determine the linear combinations with lowest entanglement and see if some kind of classical order appears in the resulting state.  

At present, interest in the harmonization of quantum and classical considerations of the same matter is also associated with the fast development of quantum technologies, including quantum computations,~\cite{King} quantum communications~\cite{Fedorov}, quantum machine learning~\cite{qlearning} and others. It means that different applied and practical problems become involved. For instance, an extreme miniaturization of classical computing devices makes inevitable the influence of quantum effects on classical computational resources. On the other hand, the creation of large-scale quantum computers ultimately requires taking into consideration the decoherence effects originating from the connection of quantum computing core and classical environment. From the algorithm perspective the solution of the  real-world computational problems such as optimization~\cite{qopt}, search~\cite{Nielsen}, image processing~\cite{qpixel} and others also aims for searching for efficient representations of classical data for following quantum processing and, conversely, converting results of quantum measurements into the appropriate classical data format. Thus, we believe that the proposed combination of the TOS  and decoherence theory can be employed to solve not only fundamental problems in condensed matter physics as it was demonstrated in this paper, but also more practical ones related to the development of quantum technologies.

\section*{Methods}
\subsection*{Exact diagonalization}
The eigenstates used for the calculation of the scalar chirality and magnetization were obtained via an exact diagonalization approach. For that purpose we used the implicitly restarted Arnoldi algorithm as implemented in ARPACK library. Such a solution allowed us to optimize memory and CPU utilization due to CRS (Compressed Row Storage) sparse matrix format~\cite{CRSMatrix} used for the representation of the Hamiltonian.  The calculation of scalar chirality as well as fidelity and local magnetization was performed on GPU using CUDA framework and cuBLAS library. We also used the package for exact diagonalization developed by Tom Westerhout \cite{Tom}. In order to construct the TOS for 19-site supercell we calculated 512 eigenstates using 2048 Arnoldi vectors for a magnetic field $B = 0.44$. For the classical TOS shown in~\ref{classic_towers}\,(c) we also calculated 16 low-lying eigenstates including the ground state using 64 Arnoldi vectors for a magnetic field in the interval $[0,1]$ with steps $\Delta{}B = 0.02$. In the case of antiferromagnets we solved the eigenproblem for each sector of the Hamiltonian separately. For each spin sector of the triangular supercell we performed a full diagonalization whereas for the square lattice supercell we calculated 512 low-lying eigenvectors.

\begin{figure}[t]
	\includegraphics[width=\columnwidth]{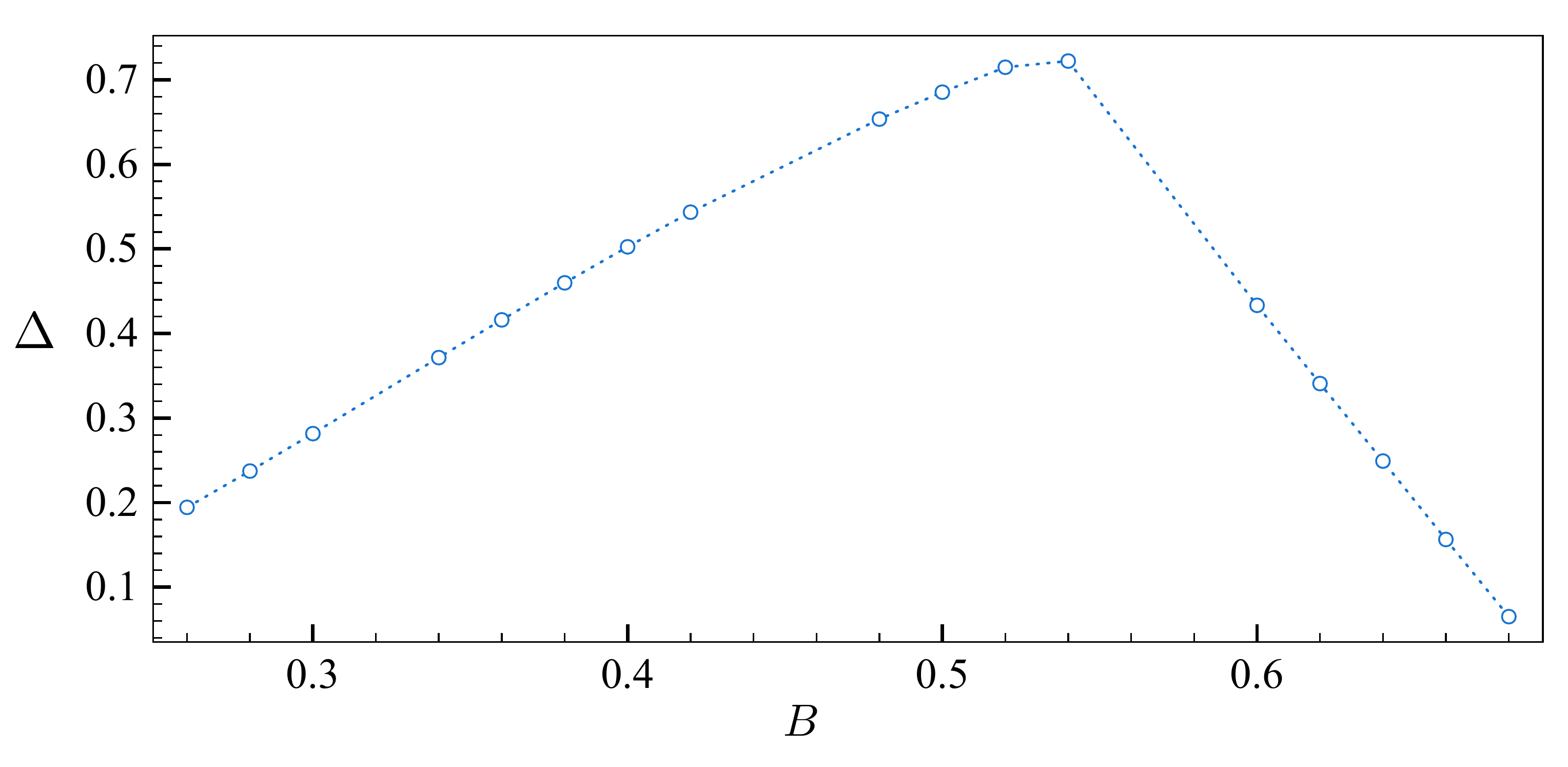}
	\caption{\label{delta} Magnetic field dependence of the energy gap between th 19 low-lying eigenstates and the rest of the spectrum as obtained in exact diagonalization calculations.}
\end{figure}

Fig.\ref{delta} shows the main feature of the eigenspectrum of the spin Hamiltonian, that is the energy gap between the low-lying eigenfunctions and the rest.  

\section*{Acknowledgements}
We would like to thank Andrey Bagrov and Tom Westerhout for useful discussions.
This work was supported by the Swiss-Russia preparation grant 2020. The work of M.I.K. was supported by the European Research Council (ERC) under the European Union's Horizon 2020 research and innovation programme, grant agreement no 854843-FASTCORR. 
The work of E.A.S. was supported by the European Union's Horizon 2020 Research and Innovation programme under the Marie Sk\l{}odowska Curie grant agreement No.~839551 - \mbox{2DMAGICS}.
Exact diagonalization calculations were performed on the Uran supercomputer at the IMM UB RAS.

\end{document}